\documentclass[10pt,article]{article}
\usepackage[top=0.85in,left=0.75in,footskip=0.75in,marginparwidth=2in]{geometry}
\usepackage{physics}
\usepackage{nameref,hyperref}
\usepackage{graphicx}
\usepackage{color}
\usepackage{abstract,lipsum}
\usepackage{cite}
\usepackage{amsmath}

\usepackage[aboveskip=1pt,labelfont=bf,labelsep=period,singlelinecheck=off]{caption}

\makeatletter
\renewcommand{\@biblabel}[1]{\quad#1.}
\makeatother

\usepackage{lastpage,fancyhdr,graphicx}
\usepackage{epstopdf}

\usepackage{wrapfig}
\usepackage[pscoord]{eso-pic}
\usepackage[fulladjust]{marginnote}
\reversemarginpar

\begin{document}
\vspace*{0.35in}

\begin{flushleft}
{\LARGE
\textbf\newline{Phase-dependent light-induced torque}
}
\newline
\\
Seyedeh Hamideh Kazemi\textsuperscript{1},
Mohammad Mahmoudi\textsuperscript{1,*}
\\
\bigskip
\textsf{1} Department of Physics, University of Zanjan, University Blvd., 45371-38791, Zanjan, Iran
\\
* mahmoudi@znu.ac.ir

\end{flushleft}
\begin{abstract}
Optical torque on individual atoms in a Bose-Einstein condensate can rotate the ensemble and generate a current flow [Phys. Rev. A 82, 051402 (2010)]. We exploit the fact to suggest a new mechanism for enhanced and well-controlled rotational motions and show how atoms, interacting with Laguerre-Gaussian beams, experience a torque whose features depend on relative phase of applied fields so that zero and large positive, or even negative, values for the torque can be obtained. Such controllable torque, along with simplicity of tuning the relative phase, can simplify a possible implementation of current flows in Bose-Einstein condensates.
\end{abstract}
\vspace*{0.2in}

\section{Introduction}

Photons, as one of the elementary particles, carry not only energy but also momentum. Like the analysis of energy, any interaction between photons and matter is inevitably accompanied by the exchange of momentum which can involve either linear \cite{Bartoli,max} or angular momentum \cite{Lifshitz}. The angular momentum carried by light can be characterized by internal or spin angular momentum (SAM) associated with the polarization of light and external or orbital angular momentum (OAM) associated with the spatial distribution of fields. The former was demonstrated experimentally by Beth \cite{beth1} in 1935 and the latter has only recently been discovered and exploited \cite{Allen}. 

For fields with simple symmetry (e.g. Laguerre-Gaussian (LG) beams), the angular momentum is associated with optical singularities \cite{Allen,Berry}. The LG modes have found use in areas such as optical micromanipulation, quantum optics, and optical communications. A considerable bulk of research has emerged dealing with LG fields and articles concerned with the study of their properties \cite{Stilgoe,Zhang}, generation \cite{Tamm,Heckenberg,Lin,Ruffato}, and uses \cite{Karimi,Otsu,Sun,Das,kazemi1,kazemi2,akin} have been actively published. For instance, three-dimensional off-axis trapping of dielectric spheres using LG beams was described in Ref.~\cite{Otsu}. Replacement of a conventional Gaussian beam with an LG one was shown to result in a reduced Doppler width in the absorption spectrum of $^{85}$Rb and $^{87}$Rb atoms \cite{Das}. Additionally, the OAM of LG beams is regarded as an additional degree of freedom for photons, thus indicating a system of a higher dimension for information transmission. Long-distance communication by means of LG beams has been recently demonstrated \cite{Krenn}. Moreover, storage and retrieval of light pulses carrying OAM have been reported, opening the way to the realization of high-dimensional quantum information networks \cite{giner}.

Exchanging momentum between photons and matter often has a mechanical consequence. In the case of the linear momentum, such mechanical effects range from comet tails to laser cooling of atoms \cite{andersen}. The transfer of SAM to atoms has been studied for over a century. In a seminal work, Poynting showed that circularly polarized light should exert a torque on a quarter-wave birefringent plate and proposed a subsequently unperformed experiment to measure it \cite{Poynting}. In an elegant extension of Poynting's idea, the torsional resonances of a fiber were exploited to detect the torque on a birefringent plate \cite{beth1}. In the past two decades, the mechanical effects of optical OAM on microscopic particles \cite{7,8,9,karen} and atoms \cite{11,12} have been investigated. For macroscopic object, SAM and OAM of light has been used to rotate it. However, transfer of SAM will not result in a mechanical rotation of an atom, transfer of the OAM can rotate the atom. In this respect, the seminal works are that of Allen and co-workers who showed that atoms moving in LG beams should exhibit novel rotational effects, in addition to expected translational ones \cite{Babiker,Power}. They showed that those effects involve changes to both internal and gross motions of the atom. As far as the gross motion is concerned, an LG beam interacting with two-level atoms creates a light-induced torque which can rotate atoms about the beam axis \cite{Babiker}. They also found that the internal motion of the atom is significantly influenced by the OAM of LG beams; Doppler shift for moving atom receive a significant additional contribution, azimuthal Doppler shift, which is directly proportional to the OAM of LG modes \cite{Power}. 

Since the first realization of atomic Bose-Einstein condensate (BEC) \cite{Cornell,Ketterle}, the physics of ultra-cold atoms has gained a great deal of attention, particularly for studying macroscopic quantum states \cite{Pethick,Brennecke,Bloch,gain,Matthews,Madison}. For instance, BEC superfluid properties can be explored by vortex state; Matthews and his colleagues \cite{Matthews} reported in 1999, the first creation of vortices in BECs through a phase-engineering scheme involving spatial and temporal control of interconversion between two components of the BEC. A year later, a BEC of $^{87}$Rb was stirred using a focused laser beam \cite{Madison}. Recently, transfer of OAM from rotating beams to atoms has been also reported \cite{andersen}. In fact, this beautiful experiment indicates two major features: an optical molasses traps atoms and the OAM exerts a torque on the center of mass of each trapped atom in such a way that the ensemble can be induced to rotate, resulting in a persistent current flow in the Bose gas. More recently, it has been shown that three-level atoms subject to two counter-propagating LG beams experience a quantized light-induced torque \cite{lembesis}. Their proposed scheme allows for controlling the rotational motion by coherent population trapping (CPT) condition. The present paper is highly motivated by the works in Refs.~\cite{andersen} and \cite{lembesis}. However, a new scheme achieving tunable and giant optical torque for controlling rotational motions in a BEC, based on phase-dependent optical effect in a closed four-level system, is suggested. 

As rightly stated by Andersen \textit{et al.} \cite{andersen}, the many-body wave-function of the whole system (a BEC of sodium atoms) can be represented by the product of single-particle wave-functions, in this paper, we will concentrate on the derivation of dissipative force acting on an atom, and show that there is a nonzero torque associated with the force, resulting in a rotational effect on the whole BEC. The system investigated is a four-level
atoms in a double-$\Lambda$ configuration, interacting with four LG beams. Note that this scheme is a closed-loop system in which atomic transitions form a closed cycle in such a way that the atomic state depends on the relative phase of applied fields. The laser configuration consists of four LG beams with same azimuthal modes, allowing for a fine control over the torque, and subsequently, rotational motion in atomic BEC. We must reiterate the importance of the fact the phase controlling of the  torque is more easier than that via the CPT condition; the main advantage of our proposed scheme over the similar schemes on rotational motions in the BEC, therefore, is its simple implementation. What is more, our model shows a wide range of tunability so that zero and very large positive (or even negative) values for the torque can be achieved. Such controllable capability, along with the simplicity, may simplify a possible implementation of controllable rotational motions in the BEC.

\begin{figure}[t]
\centerline{\includegraphics[width=9cm]{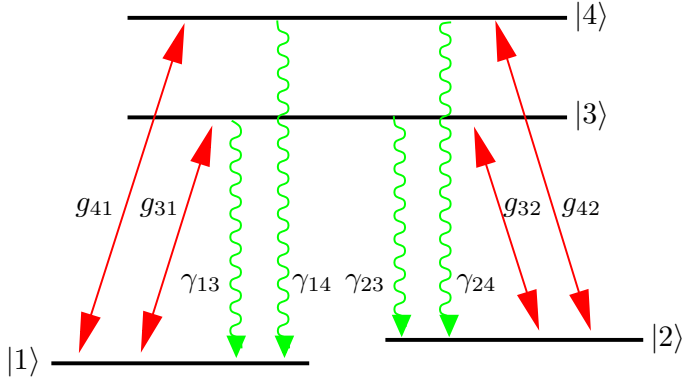}}
\vspace*{10mm}
\caption{ Energy schematic of the four-level double-$\Lambda$ atomic system. Wavy lines show spontaneous decays from the excited states.}
\label{fig1}
\end{figure}

\section{The Model}

We consider a closed-loop four-level double-$\Lambda$ atomic system interacting with four coherent fields, as shown in Fig.~\ref{fig1}. All of allowed electric dipole transitions are driven by laser fields; the excited level $\vert 3 \rangle$ is coupled to ground levels $\vert 1 \rangle$ and $\vert 2 \rangle$ by two beams with carrier frequencies $\omega_{31}$ and $\omega_{32}$, respectively. Two coherent fields with carrier frequencies $\omega_{41}$ and $\omega_{42}$, couple another excited state, $\vert 4 \rangle$, to lower states. The spontaneous decay rates on the dipole-allowed transitions are denoted by $\gamma_{13}$, $\gamma_{23}$, $\gamma_{14}$ and $\gamma_{24}$.  
The driving fields applied to transition $\vert i \rangle$ and  $\vert j \rangle$ ($i \in \lbrace3 ,4 \rbrace$ and $j \in \lbrace 1,2 \rbrace$) can be written as
\begin{equation}
\vec{E}_{ij}= \hat{e}_{ij} \, E_{ij}\, e^{-i(\omega_{ij} t -\vec{k}_{ij} \vec{r} +\phi_{ij}) } +\mathrm{c.c.},
\label{eq1}
\end{equation}
where, $\hat{e}_{ij}$, $E_{ij}$, $\vec{k}_{ij}$ and $\phi_{ij}$ are unit polarization vector, amplitude, wavevector and absolute phase, respectively. Also, c.c. denotes the complex conjugate of the first term in Eq.~(\ref{eq1}).

In dipole and rotating-wave approximation, the Hamiltonian is given by \cite{scully}
\begin{equation}
 H=  \, \sum _{m=1}^{4} E_{m} \vert m \rangle \langle m \vert -\sum_{i=3,4} \sum_{j=1,2} (\hbar g_{ij} e^{-i (\omega_{ij} t -\vec{k}_{ij} \vec{r} +\phi_{ij})} \vert i\rangle \langle j\vert +\mathrm{H.c.}),
\end{equation}
where the parameter $E_m$ denotes energy of involved states and H.c. corresponds to the Hermitian conjugate of the terms explicitly written in the Hamiltonian. Also, Rabi frequencies are defined as $g_{ij}= E_{ij} (\hat{e}_{ij}.\vec{\mu}_{ij})/ \hbar$ with $\vec{\mu}_{ij}$ being as atomic dipole moment of corresponding transitions.  

LG beams ($LG_{p}^{l}$), having a doughnut-shaped intensity distribution and zero intensity at beam center, are characterized by two indexes: radial ($p$) and azimuthal ($l$). The former is the number of radial node for radius $r>0$ and the latter, quantifying the OAM carried by the beam, represents the number of times the phase
completes 2$\pi$ on a closed loop around the axis of propagation. 
The field amplitude for the LG mode can be defined in terms of mode amplitude $\varepsilon_{klp}$ and phase factor $\Theta_{klp}$ as \cite{allen2}
\begin{equation}
E_{LG} = \varepsilon_{klp} \, e^{i \Theta_{klp}} +\mathrm{c.c.},
\end{equation}
by ignoring the $z$ dependence in the region $z\ll \,$Rayleigh length $(z_{R})$, the amplitude of the LG beam can be defined as \cite{kim}
\begin{equation}
E_{LG} =  E_{0LG} \, [\dfrac{\sqrt{2} r}{w}]^{\vert l \vert} \, L_{p}^{\vert l \vert}(\dfrac{2r^2}{w^2}) \,  \exp (  -\dfrac{r^2}{w^2} +i l \varphi )+\mathrm{c.c.},
\end{equation}
where $E_{0LG}$, $w$ and $L_{p}^{\vert l \vert}$ represent amplitude, beam waist for a beam width of $w_0$ ($w \approx w_{0} $ for $z\ll z_{R}$) and associated Laguerre polynomial \cite{allen2}. Also, ($r$, $\varphi$, $z$) are cylindrical coordinates. 

In this paper, the radial index associated with LG beams is taken as $p=0$, so the Rabi frequency for LG beams (LG$_{0}^{l}$) is written as 
\begin{equation}
g_{ij} =  g^{'}_{ij} \,(\dfrac{\sqrt{2} r}{w})^{\vert l \vert} \,  \exp (  -\dfrac{r^2}{w^2} ),
\end{equation}
with $g^{'}_{ij}$ being as the Rabi frequency constant. Note that we have assumed $p=0$, allowing only for doughnut modes of order $l$, however, the treatment can, in principle, be applied also to any field belonging to the family of LG beams. 

The density matrix equations for the system are \cite{korsun}
\begin{subequations}
\begin{align}
\frac{d}{dt}  \tilde{\rho}_{11} &= i g_{31}^{*} \tilde{\rho}_{31}- i g_{31} \tilde{\rho}_{13}+ i g_{41}^{*} \tilde{\rho}_{41} - i g_{41} \tilde{\rho}_{14} +2 \gamma_{14} \tilde{\rho}_{44}+ 2 \gamma_{13} \tilde{\rho}_{33}, \\ 
\frac{d}{dt}  \tilde{\rho}_{22} &= i g_{32}^{*} \tilde{\rho}_{32}- i g_{32} \tilde{\rho}_{23}+ i g_{42}^{*} \tilde{\rho}_{42} - i g_{42} \tilde{\rho}_{24}  + 2 \gamma_{23} \tilde{\rho}_{33}+ 2 \gamma_{24} \tilde{\rho}_{44},\\
\frac{d}{dt}  \tilde{\rho}_{33} &=- i g_{31}^{*} \tilde{\rho}_{31}+ i g_{31} \tilde{\rho}_{13}- i g_{32}^{*} \tilde{\rho}_{32} + i g_{32} \tilde{\rho}_{23} - 2 (\gamma_{13} + \gamma_{23})\tilde{\rho}_{33},\\
\frac{d}{dt}  \tilde{\rho}_{12}&= i ( \Delta_{32}-\Delta_{31} )\tilde{\rho}_{12} + i g_{31}^{*} \tilde{\rho}_{32}+ i g_{41}^{*} \tilde{\rho}_{42} e^{i \Phi}  - i g_{32} \tilde{\rho}_{13}-i g_{42} \tilde{\rho}_{14} e^{i \Phi},\\
\frac{d}{dt}  \tilde{\rho}_{13}&=- i \Delta_{31} \tilde{\rho}_{13} + i g_{31}^{*}( \tilde{\rho}_{33}- \tilde{\rho}_{11})- i g_{32}^{*} \tilde{\rho}_{12} + i g_{41}^{*} \tilde{\rho}_{43} - \gamma_{3} \tilde{\rho}_{13},\\
\frac{d}{dt}  \tilde{\rho}_{14}\ &= i(\Delta_{32}-\Delta_{31}-\Delta_{42}) \tilde{\rho}_{14}+ i g_{31}^{*} \tilde{\rho}_{34} - i g_{42}^{*}  \tilde{\rho}_{12} e^{-i \Phi}- \gamma_{4} \tilde{\rho}_{14} \\ \nonumber
 &+ i g_{41}^{*} ( \tilde{\rho}_{44}- \tilde{\rho}_{11}),\\
\frac{d}{dt}  \tilde{\rho}_{23}\ &=- i \Delta_{32} \tilde{\rho}_{23} + i g_{32}^{*}( \tilde{\rho}_{33}- \tilde{\rho}_{22})- i g_{31}^{*} \tilde{\rho}_{21} + i g_{42}^{*} \tilde{\rho}_{43} e^{-i \Phi} -\gamma_{3} \tilde{\rho}_{23}, \\
\frac{d}{dt}  \tilde{\rho}_{24}\ &=- i \Delta_{42} \tilde{\rho}_{24} + i g_{42}^{*}( \tilde{\rho}_{44}- \tilde{\rho}_{22})+i g_{32}^{*} \tilde{\rho}_{34} e^{i \Phi} - i g_{41}^{*} \tilde{\rho}_{21} e^{i\Phi}-\gamma_{4} \tilde{\rho}_{24},\\
\frac{d}{dt}  \tilde{\rho}_{34}\ &=- i(\Delta_{42}-\Delta_{32}) \tilde{\rho}_{34} + i g_{31}  \tilde{\rho}_{14} +i g_{32} \tilde{\rho}_{24} e^{-i \Phi} - i g_{41}^{*} \tilde{\rho}_{31}  - i g_{42}^{*} \tilde{\rho}_{32} e^{-i \Phi}\\ \nonumber
&- (\gamma_3+\gamma_4) \tilde{\rho}_{34}.
\end{align}
\end{subequations}
The remaining equations follow from constraints $\tilde{\rho}_{mn}=\tilde{\rho}^{*}_{nm} \,$ $(m,n$ $ \in \lbrace1, ..., 4\rbrace)$ and $\sum _{m} \tilde{\rho}_{mm}$ $=1 $. The parameter $\rho_{mn} = \vert m\rangle \langle n \vert$ and $\tilde{\rho}_{mn}\,$ $= \rho_{mn} \, \exp[i(\omega_{mn} t -k_{mn} z - l_{mn} \varphi +\phi_{mn})]$ denotes the corresponding operator in the new reference frame. Also, $\Delta_{ij}= \omega_{ij}-\bar{\omega}_{ij}$ and  $\bar{\omega}_{ij}= (E_{i}-E_{j})/ \hbar$ represent the detuning of the laser field from corresponding transition and transition frequency, respectively. We have further defined $ \gamma_i= \gamma_{1i}+ \gamma_{2i}$. The parameter $\Phi $ is given by
\begin{subequations}
\begin{align}
\Phi &= \Delta t- \vec{K} \vec{r}-L \varphi +\phi_0,  \\
\Delta &= (\Delta_{32} +\Delta_{41})-(\Delta_{31} +\Delta_{42}),  \\
\vec{K} &= (\vec{k}_{32}+\vec{k}_{41})-(\vec{k}_{31}+\vec{k}_{42}), \\
L&= (l_{32}+l_{41})-(l_{31}+l_{42}), \\
\phi_{0} &= (\phi_{32}+\phi_{41})-(\phi_{31}+\phi_{42}).
\label{eq2}
\end{align}
\end{subequations}
The parameters $\Delta$, $\vec{K}$, $L$, and $\phi_{0} $ are the multiphoton resonance detuning, wavevector mismatch, azimuthal index difference, and initial phase difference, respectively. Noting that the system does not have a constant steady-state solution, however, in multiphoton resonance condition, \textit{i.e.,} $\Delta=0$ and $\vec{K}=0$, a stationary solution in the long-time limit can be found so that it can be determined by the relative phase $\phi_{0}$.

We then proceed to derive the formula for dissipative force on the atom in LG beams. Following the same procedure as for the case of a two-level system in a plane wave \cite{cook}, the expression for dissipative force, also called radiation pressure or scattering force, is given in terms of off-diagonal density-matrix elements and the phase factor associated with beams ($ \vartheta_{ij} =l_{ij} \varphi + k_{ij} z$ )
\begin{align}
F_{diss}  &=  -i \hbar \, \sum_{i=3,4} \sum_{j=1,2}   \, \nabla \vartheta_{ij}  \, (\rho_{ij} + \rho_{ji}) \, [g_{ij} e^{-i(\omega_{ij} t -k_{ij} z - l_{ij} \varphi +\phi_{ij})} \\ \nonumber
&- g_{ij}^{*} e^{i(\omega_{ij} t -\vec{k}_{ij} \vec{r} - l_{ij} \varphi +\phi_{ij})}]= i \hbar  \, \nabla \vartheta_{ij}  \, (g_{ij} \, \tilde{\rho}_{ij} - g_{ij}^{*} \, \tilde{\rho}_{ji}).
\end{align}
It worth mentioning that radiation force is calculated within semi-classical approximation, in which the spatial extent of the atomic wavepacket is assumed to be much smaller than the optical wavelength so that atomic position operator can be replaced by its expectation value \cite{minogin}. As is seen, the radiation force can be obtained by substituting the solutions for $\rho_{ij}$ into Eq.~(8) and again discarding terms that oscillate at twice the optical frequency.

As the force depends on the gradient in cylindrical coordinates, $\nabla \vartheta_{ij}=l_{ij}/r\, \hat{\varphi} + k_{ij} \hat{z}$, we have nonzero force component in the direction $\hat{\varphi}$. We then assume that all laser fields have the same azimuthal indexes ($l_{ij}=l$) and the dissipative force takes the form
\begin{align}
F_{diss} &=\dfrac{ \hbar l}{r} \sum_{i=3,4} \, \sum_{j=1,2} \sigma_{ij} \hat{\varphi}+   \hbar \sum_{i=3,4} \sum_{j=1,2} k_{ij}\, \sigma_{ij} \hat{z},
\end{align}
with
\begin{align}
\sigma_{ij} &=  i(g_{ij} \, \tilde{\rho}_{ij} - g_{ij}^{*} \, \tilde{\rho}_{ji}).
\end{align}
It is clear that appeared azimuthal component of the dissipative force is responsible for a torque acting on the atom about the beam axis, whose explicit expression can be written as follows
\begin{equation}
T_{diss}=  r \times F_{diss}  =  \hbar l \sum_{i=3,4} \, \sum_{j=1,2} \sigma_{ij} \hat{z}.
\end{equation}

Indeed, such feature is completely predictable from results already derived for the case of a plane wave interacting with an atom in which the dissipative force would be proportional to the wavevector. In fact, any azimuthal fileds, e.g., LG ones, acting on the atom is expected to have associated with a wavevector, thus, it is reasonable to expect that such azimuthal force should exert a torque on the atom \cite{Babiker}. Noting that this torque is quantized, due to its explicit proportionality to the integer $l$. Evidently just as the OAM can change the force and its accompanying torque, changes in the function $\sum_{i} \sum_{j} \sigma_{ij}= \sigma$ (after that, we call it "torque function"), will change them.

\section{Results}
\begin{figure*}[!ht]
\centerline{\includegraphics[width=12cm]{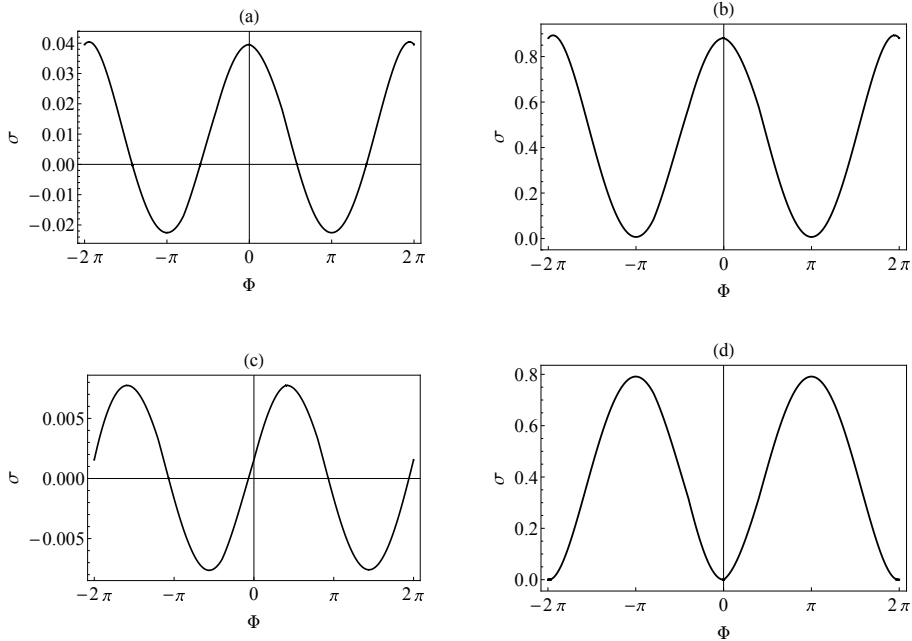}}
\vspace*{10mm}
\caption{The torque function ($\sigma$), directly describing the torque, in units of $s^{-1}$ versus the relative phase of applied fields ($\Phi$) for following parameters: (a) $\Delta_0=0$, $g^{'}_{31}=g^{'}_{42}=0.1 $, and $ g^{'}_{32}= g^{'}_{41}=1$, (b)  $g^{'}_{32}= g^{'}_{41}=5 $ and with other parameters same as those defined previously, (c) with same Rabi frequencies in (a) and $\Delta_0=5 $, (d) $g^{'}_{32}= g^{'}_{41}= g^{'}_{31}=g^{'}_{42}=5 $ and $\Delta_0=0$. }
\label{fig2}
\end{figure*}

As mentioned before, in the previously proposed scheme for light-induced torque \cite{lembesis}, three-level atoms experience a torque whose features depend on the detuning of applied fields so that the rotational motion in the BEC can be changed by tuning the resonance condition between fields and atomic transitions. Keeping in mind their innovative ideas, the main question is now: Is there any way to improve on the scheme to bring about a more effective control over such effects? In the following, we show how our suggested scheme results in an easy control over the torque without encountering difficulty in changing detunings as well as in an enhanced torque. It is imperative to note here that a very similar scheme to that presented in Fig.~\ref{fig1}, was used in an experimental work to show how phase dependence can be used to control optical properties of an atomic system \cite{huss2}. This seminal work on a closed-loop double-$\Lambda$ atomic system, described two effective techniques for variation of the relative phase of applied fields; a mechanical phase shifter and an electro-optical phase modulator. Further, it was emphasized that both techniques, after an appropriate calibration, can allow direct adjustment of the relative phase. Thus, from the experimental point of view, the phase controlling is much easier than changing the resonance condition.

Before presenting numerical results for the dissipative force, it is desirable to point out some considerations: As a realistic example, we consider the hyperfine energy levels of $^{85}$Rb \cite{steck}. Beam orientation should be determined by criteria of the mismatch condition and detunings, throughout the paper, are determined by the criteria of multiphoton resonance condition. All calculations are performed by assuming the beam width of each LG beam as $w= 1\,$mm, azimuthal indexes as $l=1$, and equal spontaneous decay rates, $ \gamma_{13}= \gamma_{14}=  \gamma_{23}= \gamma_{24}=\gamma$ \cite{Turnbull}. Furthermore, our results are represented in scaled quantities to obtain best possible comparison with other similar schemes; torque function is represented in unit of $s^{-1}$. As is conventional, light propagation problems are considered in terms of dimensionless variables; dimensionless Rabi frequency constants (field amplitude) $ g^{'}_{ij}/\gamma$ and detunings $\Delta_{ij} / \gamma$. Additionally, we assume as in Ref.~\cite{andersen} that each atom will receive an identical light-induced torque, and thus, rotational effects on the whole BEC are expected due to this torque on individual atoms. Recalling that the many-body wave-function of the BEC is represented by the product of identical single-particle wave-functions.


We start our analysis with the case of equal detunings, $\Delta_{31}=\Delta_{32}=\Delta_{41}=\Delta_{42}=\Delta_{0}$. In Fig.~\ref{fig2}, we depicts the variation of the torque function ($\sigma$) versus the relative phase of applied fields for following values of parameters: (a) $\Delta_0=0$, $g^{'}_{31}=g^{'}_{42}=0.1 $, $ g^{'}_{32}= g^{'}_{41}=1$, (b) $ g^{'}_{32}= g^{'}_{41}=5 $ and with other parameters same as those defined above, (c) with same Rabi frequencies in Fig.~\ref{fig2}(a) and $\Delta_0=5 $, (d) $g^{'}_{31}=g^{'}_{32}= g^{'}_{41}=g^{'}_{42}=5 $ and $\Delta_0=0$. First, we will consider the case of a complete one-photon resonance, $\Delta_{ij}=0$, in which atoms are pumped by one of the resonant laser pairs ($E_{32}$ and $E_{41}$). Fig.~\ref{fig2}(a) demonstrates the phase dependence of the torque function in units of $s^{-1}$. From what has been just discussed, it is clear that the optical properties of the system depend on the relative phase of the fields: $\Phi =\phi_{0}=\phi_{32}+\phi_{41}-(\phi_{31}+\phi_{42})$, due to the closed-loop configuration. We see that at phases around $\Phi=\pm  \pi/2$ and $\pm 3 \pi /2$, the torque vanishes, while near $\Phi=0$ and $\pm 2 \pi$ the torque reach maxima. As  azimuthal indexes are assumed to be positive, negative torque opposite to the handedness of LG fields is illustrated by the negative sign of the torque function. It is clear that, by modulating the relative phase of applied fields, we can obtain zero and also positive (or even negative) values for the torque function and achieve controlling over the behavior of the torque, correspondingly. Fig.~\ref{fig2}(b) depicts corresponding variation of the torque function expressed in the same unit, when Rabi frequencies of pump fields, $E_{32}$ and $E_{41}$, are increased to 5. The other parameters are kept the same as in Fig.~\ref{fig2}(a). Somewhat similar results hold for this case; maximum appears again around previous values of the relative phase, $\Phi=0$ and $\pm 2 \pi$, but larger torque in the maxima. However, the torque function appears to approach zero around $\Phi=\pm  \pi$. Since such double-$\Lambda$ system with resonant excitation and with moderate Rabi frequencies, a few $\gamma$, was used in an experimental works in Ref.~\cite{huss2}, it then seems reasonable to suggest that a further enhancement of well-controlled torque could be easily obtained for those parameters. Fig.~\ref{fig2}(c), on the other hand, shows the torque function for same Rabi frequencies in Fig.~\ref{fig2}(a) and $\Delta_0=5$. A somewhat different trend is seen for non-resonant excitation:  Smaller torque at maxima and zeros at $\Phi=-2 \pi/30 , -2 \pi/30 + n \pi$ and  $-(2 \pi/30 + n \pi)$ with $n$ as an integer number. The torque function for resonant excitation with larger equal-Rabi frequency fields, $g^{'}_{31}=g^{'}_{32}= g^{'}_{41}=g^{'}_{42}=5 $, is shown in Fig.~\ref{fig2}(d). The general trend is the same as that displayed in Fig.~\ref{fig2}(b) in which the function oscillates between zero and a large positive maximum.

It may be mentioned that here we present results for incident LG doughnut modes of $l=1$, however, the torque is proportional to azimuthal index ($l$) and, consequently, is expected to be increased by using LG doughnut modes with large values of $l$. In this context, it would be imperative to mention that winding numbers ($l$) as large as 300 can be experimentally achieved \cite{fick}. As mentioned previously, light-induced torque on individual atoms will have a rotational effect on the whole BEC, thus, this scheme with such controllable capability, along with simplicity of the tuning the relative phase \cite{huss1,huss2}, appears to be very promising for having a controllable rotational motion in the BEC.

We then proceed to investigate another double-$\Lambda$ configuration which has been used in recent experimental works on atomic vapor \cite{qin,Turnbull}; one of the laser pairs ($E_{31}$ and $E_{32}$) is resonant with their correspondence transitions, i.e., $\Delta_{31}=\Delta_{32}=0$, while the other pair is far detuned from the excited state $\vert 4 \rangle$: $\Delta_{41}=\Delta_{42}=\Delta_{1}$. Numerical results show that the radiation force oscillates between two very small values (nearly zero), which indicates that torque acting on those atoms will approach zero. In order to further analyze such behavior, we proceed to investigate analytical solution for off-diagonal density-matrix elements. By solving equations of motion in the steady-state situation, which can be obtained from Eqs.~(6) for vanishing time derivatives, we can derive analytical formula of off-diagonal density-matrix elements ($\tilde{\rho}_{ij}$), and consequently, the torque function. For the case of large detunings- $\Delta_{1} \gg \gamma,$ $\, g^{'}_{ij}, \, g^{'}_{ij} g^{'}_{i^{'}j^{'}}$- the
expressions for $\tilde{\rho}_{ij}$ is given as \cite{korsun}

\begin{align}
&\mathrm{v}_{31}  = -\dfrac{g_{32} g_{41} g_{42}}{g_{0}^{2} \Delta_{1}} \sin \Phi,  \nonumber \\
&\mathrm{v}_{32}  = \dfrac{g_{31} g_{41} g_{42}}{g_{0}^{2} \Delta_{1}} \sin \Phi, \nonumber \\
&\mathrm{v}_{41}  = \dfrac{g_{31} g_{32} g_{42}}{g_{0}^{2} \Delta_{1}} \sin \Phi, \nonumber \\
&\mathrm{v}_{42}  =-\dfrac{g_{31} g_{32} g_{41}}{g_{0}^{2} \Delta_{1}} \sin \Phi,
\label{eq12}
\end{align}
where $\mathrm{v}_{ij}=\mathrm{Im}[\tilde{\rho}_{ij}]$ and $g_{0}^{2}= g_{31}^{2} + g_{32}^{2}$. Noting that terms
of our interest would be the imaginary parts of $\tilde{\rho}_{ij}$, as $\sigma_{ij}$ is proportional to $\mathrm{Im}[\tilde{\rho}_{ij}] $. A careful analysis of Eq.~(\ref{eq12}) reveals that- for all values of $g_{ij}$, whether equal or not, and that of $\Delta_{1}$- the torque function would be zero, as two terms in the function ($\sigma_{32}$ and $\sigma_{41}$) eliminate other ones ($\sigma_{31}$ and $\sigma_{42}$).

As final notes, it is important to highlight key advantages of the suggested scheme. Our approach permits an easy implementation of the enhanced torque, as the relative phase can be easily changed. Further advantage of approach we have used is that it shows a wide range of tunability so that zero and positive, or even negative, values for the torque can be obtained just only by modulating the relative phase. In addition, the torque can be enhanced substantially, compared with the similar work in which the maximum value of the function was about $0.055$ \cite{andersen}. From what is discussed above, it can be concluded that these enhanced and well-controlled rotational effects in the BEC should be amenable to experimental realization, because it is straightforward to generate LG  beams with large azimuthal modes and to adjust the relative phase of applied fields in the laboratory. On the basis of our results, we also envisage the possibility of generating current flows in atomic gas BECs, more effective and less effortful than existing methods. 


\section{Conclusions}

We have investigated the dissipative force on double-$\Lambda$ four-level atoms in a laser configuration consisting of four LG beams with same azimuthal indexes. Our analysis has shown that, in addition to usual dissipative force, the atom experiences a torque, arising from the OAM properties of LG beams, whose features sensitively depend on the relative phase of applied fields. Under multi-photon resonance condition and for the parameters explored in this work, we can obtain zero and large positive, or even negative, values for the torque only by modulating the relative phase. Such controllable capability, simplicity of the tuning the relative phase as well as the enhanced torque, can lead to the suggestion of well-controlled rotational motions in the BEC.

\section*{Funding Information}
Iran National Science Foundation (INSF) (96008805).

\bibliographystyle{}

\end{document}